\documentclass[usenatbib]{mn2e}
\usepackage{amsmath, amssymb}
\usepackage{amsfonts}
\usepackage{epsfig,rotating}

\def\simeq{
\mathrel{\raise.3ex\hbox{$\sim$}\mkern-14mu\lower0.4ex\hbox{$-$}}
}

\def\ltsima{$\; \buildrel < \over \sim \;$}
\def\simlt{\lower.5ex\hbox{\ltsima}}
\def\gtsima{$\; \buildrel > \over \sim \;$}
\def\simgt{\lower.5ex\hbox{\gtsima}}

\def\msun{{\rm M_{\odot}}}

\def\be{\begin{equation}}
\def\ee{\end{equation}}

\def\del#1{{}}
\def\ltsima{$\; \buildrel < \over \sim \;$}
\def\simlt{\lower.5ex\hbox{\ltsima}}
\def\gtsima{$\; \buildrel > \over \sim \;$}
\def\simgt{\lower.5ex\hbox{\gtsima}}

\newcommand{\apj}{ApJ}

\newcommand{\mnras}{MNRAS}
\newcommand{\aap}{A\&A}
\newcommand{\araa}{ARA\&A}
\newcommand{\apjl}{ApJL}
\newcommand{\aj}{AJ}
\newcommand{\nat}{Nature}

\title[The small observed scale of AGN--driven outflows]{The small observed scale of AGN--driven outflows, and inside--out disc quenching}
\author[Kastytis Zubovas and Andrew King]{Kastytis Zubovas$^{1,\star}$ and Andrew King$^{2,3,4}$ \\
  $^{1}$Center for Physical Sciences and Technology, Savanori\c{u} 231, Vilnius LT-02300, Lithuania \\
  $^{2}$Department of Physics \& Astronomy, University of Leicester, Leicester, LE1 7RH, UK \\
  $^{3}$ Astronomical Institute Anton Pannekoek, University of Amsterdam, Science Park 904, 1098 XH Amsterdam, Netherlands\\ 
$^4$ Leiden Observatory, Leiden University, Niels Bohrweg 2, NL-2333 CA Leiden, Netherlands \\
  $^{\star}$ {E-mail:~} {\rm kastytis.zubovas@ftmc.lt} }

\begin{document}

\maketitle

\begin{abstract}
  Observations of massive outflows with detectable central AGN
  typically find them within radii $\lesssim 10$\,kpc. We show that
  this apparent size restriction is a natural result of AGN driving if
  this process injects total energy only of order the gas binding
  energy to the outflow, and the AGN varies over time (`flickers') as
  suggested in recent work.  After the end of all AGN activity the
  outflow continues to expand to larger radii, powered by the thermal
  expansion of the remnant shocked AGN wind.  We suggest that on
  average, outflows should be detected further from the nucleus in
  more massive galaxies. In massive gas--rich galaxies these could be
  several tens of kpc in radius. We also consider the effect that
  pressure of such outflows has on a galaxy disc. In moderately
  gas--rich discs, with gas-to-baryon fraction $< 0.2$, the outflow
  may induce star formation significant enough to be distinguished
  from quiescent by an apparently different normalisation of the
  Kennicutt-Schmidt law. The star formation enhancement is probably
  stronger in the outskirts of galaxy discs, so coasting outflows
  might be detected by their effects upon the disc even after the
  driving AGN has shut off. We compare our results to the recent
  inference of inside--out quenching of star formation in galaxy
  discs.
  
\end{abstract}


\section{Introduction}

Modern galaxy evolution models typically include feedback from active
galactic nuclei (AGN) in order to explain the drop-off in the galaxy
mass function compared with the expected halo mass function above $M_*
\simeq 10^{11} \msun$, prevent the cooling catastrophe in galaxy
clusters and produce the scaling relations between galaxies and their
central supermassive black holes (SMBH). The existence of a feedback
link has been all but confirmed observationally, with pc-scale
relativistic winds \citep[e.g.,][]{Tombesi2010A&A,Tombesi2010ApJ} and
massive outflows on scales from sub-kpc \citep{Alatalo2011ApJ} to
several kpc \citep{Feruglio2010A&A, Sturm2011ApJ, Rupke2011ApJ,
  Cicone2014A&A} detected in a number of galaxies, sometimes both
types seen in the same object \citep{Tombesi2015Natur}.

Despite this robust general picture, a number of questions remain
regarding the effects of AGN feedback upon the host galaxies. One
question is the range of spatial scales over which outflows are
found. Simple models \citep[e.g.,][]{Zubovas2012ApJ} predict outflows
propagating with roughly constant velocity out to very large
radii. However, AGN outflows are only detected within the central
$\sim 10$~kpc of the nucleus \citep[e.g.,][]{Spence2016MNRAS}. The
fact that AGN outflows should expand for a long time after the driving
AGN switches off \citep[e.g.,][]{King2011MNRAS} might offer an
explanation to this problem.

Another question is the possibility of the outflow triggering star
formation in the galaxy disc. Such a process has been proposed and
investigated before \citep{Silk2005MNRAS, Silk2009ApJ,
  Gaibler2012MNRAS, Zubovas2013MNRASb, Silk2013ApJ, Bieri2016MNRAS},
but there is still no consensus. Observations do not provide a unique
answer either, with AGN activity found associated with both elevated
\citep[e.g.,][]{Santini2012A&A, Bernhard2016MNRAS} and suppressed
\citep[e.g.,][]{Page2012Natur, Carniani2016arXiv} star formation in
the host galaxy, while sometimes no connection is evident at all
\citep[e.g.,][]{Bongiorno2012MNRAS, Mullaney2012MNRAS}. One
interesting piece of evidence is recent detection of star formation
being quenched from the inside out in galaxy discs
\citep{Tacchella2015Sci}. We suggest that similar behaviour may be
expected from disc galaxies affected by AGN outflows.

In this paper, we investigate both the propagation of AGN-driven
outflows and their effect on galaxy discs by means of a
semi-analytical model. We track the expansion of an outflow driven by
a flickering AGN by numerically integrating the equation of motion and
find that the outflow is more likely to be detected close to the
centre of the galaxy than far away. We assume that the growth of the
central black hole and the accompanying AGN activity switches off once
the total energy injected into the outflow is of order the binding
energy of the gas, because this starves the central black hole. With
this assumption we show that outflows are only detectable within $\sim
10$~kpc of the centre of the host galaxy while the galaxy nucleus is
active. Next we use a simple prescription based on the KS law to
calculate the expected star formation rate and dynamical pressure in
the galactic disc and compare this with the outflow pressure. We find
that AGN outflows can produce a significant enhancement of star
formation, especially in the outskirts of galaxy discs. This process
helps eventually quench star formation from inside out, similar to the
suggestion in recent work \citep{Tacchella2015Sci}, by exhausting the
available gas supply. It offers a diagnostic tool for detecting
remnant outflows which might be unobservable directly.

We structure this paper as follows. In Section \ref{sec:windmodel}, we
briefly describe the basics of the AGN wind outflow model. In Section
\ref{sec:expansion}, we describe the numerical integrator used to
follow the evolution of the outflow driven by a flickering AGN and
present the results of outflow size. In Section \ref{sec:burst}, we
consider the effect of such an outflow upon the galactic disc and
estimate the strength of the starburst. We discuss in Section
\ref{sec:discuss} and conclude in Section \ref{sec:concl}.

\section{Wind outflow model}\label{sec:windmodel}

To model the AGN feedback we consider the interaction of AGN winds
with the surrounding medium, which produces large--scale
outflows. This has been presented and investigated in various papers,
both analytical \citep{King2003ApJ, King2005ApJ, King2010MNRASa,
  Zubovas2012ApJ, Faucher2012MNRASb}, and numerical
\citep{Nayakshin2010MNRAS}.  Here we briefly summarize the main
features. For an extensive summary see \citet{King2015ARA&A}.

The central supermassive black hole in a galaxy accretes material from
the surroundings via a geometrically thin, optically thick disc. At
sufficiently high accretion rates, radiation pressure from the disc
and corona drives a wind with mass outflow rate comparable to the
accretion rate, i.e. $\dot{M}_{\rm w} \simeq \dot{M}_{\rm acc}$. The
wind is launched with a velocity of order the escape velocity at the
launch radius, $v_{\rm w} \simeq \eta c$, with $\eta \simeq 0.1$ the
radiative efficiency of accretion. The wind self--regulates to stay at
a Compton optical depth of approximately unity, so that each photon of
the AGN radiation field scatters on average once before escaping from
the wind. This leads to the wind momentum rate
\begin{equation}
\dot{P}_{\rm w} = \dot{M}_{\rm w} v_{\rm w} \simeq \frac{L_{\rm AGN}}{c} = \dot{P}_{\rm AGN},
\label{mom}
\end{equation}
and an energy rate
\begin{equation}
\dot{E}_{\rm w} = \frac{\dot{M}_{\rm w} v_{\rm w}^2}{2} \simeq \frac{\eta L_{\rm AGN}}{2}.
\label{en}
\end{equation}

The wind collides with the interstellar medium (ISM) surrounding the
AGN and shocks against it, driving an outflow. The dynamics of the
outflow depend on whether the shocked wind cools efficiently. In this
paper we are interested in the large--scale outflows that follow the
establishing of the $M -\sigma$ relation and sweep the galaxy clear of
interstellar gas. In this phase all cooling processes are negligible,
so the wind shock is adiabatic.  We call this type of outflow
energy--driven, because all the AGN wind energy is used to drive the
outflow \citep[but note that the forward shock into the ISM need not
  be adiabatic -- in fact there are indications that it may cool
  strongly; see, e.g.,][]{Zubovas2014MNRASa, Nayakshin2012MNRASb}.
Under ideal circumstances, the AGN wind has a kinetic energy rate
$\dot{E}_{\rm out} = \dot{E}_{\rm w}$ and a momentum rate
$\dot{P}_{\rm out} \sim 20 \dot{P}_{\rm w}$
\citep{Zubovas2012ApJ}. These winds drive the interstellar gas out of
the galaxy bulge at rates of several $100\msun\,{\rm yr}^{-1}$ and
speeds $\sim 1000\,{\rm km s}^{-1}$ These results are remarkably
consistent with observations of large--scale molecular outflows in
galaxies \citep{Rupke2011ApJ, Feruglio2010A&A, Sturm2011ApJ,
  Riffel2011MNRAS, Riffel2011MNRASb, Cicone2014A&A}, and in one case a
driving AGN wind with parameters very close to eqs. (\ref{mom},
\ref{en}) is seen simultaneously \citep{Tombesi2015Natur}.

Despite these successes, some parts of this picture need
clarification. In particular, most outflows are observed at distances
$\lesssim$ 10 kpc from the nucleus of the galaxy
\citep[e.g.,][]{Shih2010ApJ, Garcia-Burillo2015A&A,
  Spence2016MNRAS}. The analytical models predict outflow velocities
$v_{\rm out} \simeq 1000\,{\rm km s}^{-1} \simeq 1\,{\rm kpc
  Myr}^{-1}$ when the AGN is active. Naively, this would mean that
observations generally select outflows no more than a few Myr after
the start of AGN activity, which is at first sight puzzling. But there
is considerable evidence that AGN activity is very time--dependent,
with only a short duty cycle \citep{Schawinski2015MNRAS,
  King2015MNRAS}. In the next Section we argue that outflows driven by
such `flickering' AGN explain this and other features of the observed
outflows.

\section{Outflow expansion}\label{sec:expansion}

\subsection{Outflow equation of motion}

An energy-driven spherically symmetric outflow expands because of the
adiabatic expansion of the wind bubble inside it. The equation of
motion for the contact discontinuity between the shocked wind and the
shocked ISM can in principle be derived \citep[cf.][]{King2005ApJ,
  Zubovas2012MNRASb} from Newton's second law and the energy
equation. In previous papers we derived this equation of motion for
outflows with several simplifying assumptions. Here, we derive it in
the most general case which is analytically tractable, keeping only three
assumptions:
\begin{itemize}
\item the whole system is always spherically symmetric;

\item the shocked wind is completely adiabatic;

\item the wind bubble moves as a single entity, i.e. there are
  no significant radial velocity gradients inside the shocked wind region
 because the sound crossing time is much shorter than other timescales relevant in the problem.
\end{itemize}

With these assumptions the motion of the shocked wind bubble is
completely specified in terms of the radius $R(t)$ of the contact
discontinuity with the ISM.  This allows a very simple description at
the cost of some loss of detail, since we do not specify the physics
of the forward shock, and so exactly how mass is swept up by the
expanding wind bubble. Fortunately we will see that our conclusions
are not sensitive to the various assumptions one can make here. We
consider the effects of making different assumptions in Section
\ref{sec:detect}.

In deriving the equation of motion, we specify all vector quantities
with respect to the distant upstream medium. We start with the
appropriate expression of Newton's second law:
\begin{equation}
\frac{{\rm d}}{{\rm d}t}\left[M\dot{R}\right] = 4\pi R^2 P - \frac{GM\left[M_{\rm b}+M/2\right]}{R^2}.
\end{equation}
Here the term on the left hand side is the time derivative of the
linear momentum, with $M(R)$ the instantaneous swept--up gas mass being driven out when
the bubble (contact discontinuity) is at radius $R$,
$P$ is the expanding gas pressure and $M_{\rm b}$ is the mass of the stars and dark matter
within $R$ (these are left unmoved by the outflow). The factor $1/2$ in the
gravity term arises since gas is interacting with
itself. This equation gives the gas pressure as
\begin{equation} \label{eq:pres_cd}
P = \left(4\pi R^2\right)^{-1}\left[\dot{M}\dot{R} + M\ddot{R} + \frac{GM\left(M_{\rm b}+M/2\right)}{R^2}\right].
\end{equation}

Next, we take the energy equation:
\begin{equation}
\frac{{\rm d}}{{\rm d}t}\left[\frac{3}{2}PV\right] = \frac{\eta}{2}L_{\rm AGN} - P\frac{{\rm d}V}{{\rm d}t} - \frac{{\rm d}E_{\rm g}}{{\rm d}t},
\end{equation}
where the term on the left hand side is the change in wind internal
energy, $V$ is the volume cleared by the outflowing gas and $E_{\rm
  g}$ is the gravitational binding energy of the gas. We neglect 
the kinetic energy of the shocked wind gas since the wind shock is very strong
($v_{\rm w} >> \dot{R}$). The gravitational binding energy change is
\begin{equation}
\begin{split}
  \frac{{\rm d}E_{\rm g}}{{\rm d}t} &= -\frac{{\rm d}}{{\rm d}t} \left[\frac{GM\left(M_{\rm b} + M/2\right)}{R}\right] \\&= -G\left[\frac{\dot{M}M_{\rm b} + M\dot{M}_{\rm b} + M\dot{M}}{R} - \frac{\dot{R}M\left(M_{\rm b} + M/2\right)}{R^2}\right].
\end{split}
\end{equation}
The first term gives the increase in
gravitational binding energy through the increase in outflowing and
background mass, while the second gives the decrease in binding
energy from the outflow expansion (the binding energy per
unit outflowing mass decreases with $R$, as expected).

The change in internal energy depends to some extent on how the swept--up ISM
is added to the outflow. Because the shocked wind gas is extremely dilute, the contact
discontinuity is highly Rayleigh--Taylor unstable \citep{King2005ApJ, King2010MNRASb} so the shocked ISM is mixed
with it rather than forming a thin shell ahead of it. 
But the fact that the observed outflows show molecular gas moving at speeds $\sim 
1000\,{\rm km\, s^{-1}}$ probably also means that the interstellar gas cools very rapidly. By keeping the assumptions of spherical symmetry and outflow uniformity as specified in the beginning of this Section, we find
\begin{equation}
\begin{split}
  \frac{{\rm d}}{{\rm d}t}\left[\frac{3}{2}PV\right] &
  = \frac{{\rm d}}{{\rm d}t}\left[2\pi R^3P\right] \\ &= \frac{{\rm d}}{{\rm d}t}\left[\frac{\dot{M}R\dot{R}+MR\ddot{R}}{2}+\frac{GM\left(M_{\rm b}+M/2\right)}{2R}\right] \\ &= \frac{\ddot{M}R\dot{R}+\dot{M}\dot{R}^2+\dot{M}R\ddot{R}+\dot{M}R\ddot{R}+M\dot{R}\ddot{R}+MR\dddot{R}}{2} \\ &+\frac{G}{2}\left[\frac{\dot{M}M_{\rm b} + M\dot{M}_{\rm b} + M\dot{M}}{R} - \frac{\dot{R}\left(MM_{\rm b} + M^2/2\right)}{R^2}\right].
\end{split}
\end{equation}

The $P$d$V$ work term is
\begin{equation}
P\frac{{\rm dV}}{{\rm d}t} = 4\pi R^2 P \dot{R} = \dot{M}\dot{R}^2 + M\dot{R}\ddot{R} + \frac{G \dot{R}\left(MM_{\rm b} + M^2/2\right)}{R^2}.
\end{equation}

Combining the above expressions leads to
\begin{equation}
\begin{split}
\frac{\eta}{2}L_{\rm AGN} &= \dot{M}R\ddot{R} + \frac{3}{2}\dot{M}\dot{R}^2 + \frac{3}{2}M\dot{R}\ddot{R} + \frac{1}{2}\ddot{M}R\dot{R} + \frac{1}{2}MR\dddot{R} \\ &- \frac{G}{2}\left[\frac{\dot{M}M_{\rm b} + M\dot{M}_{\rm b} + M\dot{M}}{R} + \frac{3\dot{R}\left(MM_{\rm b} + M^2/2\right)}{R^2}\right].
\end{split}
\end{equation}
We rearrange this equation as
\begin{equation} \label{eq:eom}
  \begin{split}
    \dddot{R} &= \frac{\eta L_{\rm AGN}}{M R} - \frac{2\dot{M} \ddot{R}}{M} - \frac{3\dot{M} \dot{R}^2}{M R} - \frac{3\dot{R} \ddot{R}}{R} - \frac{\ddot{M} \dot{R}}{M} \\ & +\frac{G}{R^2}\left[\dot{M} + \dot{M}_{\rm b} + \dot{M}\frac{M_{\rm b}}{M} - \frac{3}{2}\left(2M_{\rm b}+M\right)\frac{\dot{R}}{R}\right].
  \end{split}
\end{equation}
In all the equations above, $\dot{M}\equiv \dot{R}\partial M/\partial
R$ and $\ddot{M} \equiv \ddot{R}\partial M/\partial R + \dot{R}
({\rm d}/{\rm d}t)\left(\partial M/\partial R\right)$. In the
rest of the paper, we refer to the first term on the right hand side
of equation (\ref{eq:eom}) as the driving term, the next four terms as the
kinetic terms, and the final term (involving the square brackets) as the gravity term.

\subsection{Numerical model}

We integrate equation (\ref{eq:eom}) numerically to follow the outflow
expansion for a specified history of nuclear activity, expressed as a
function $L_{\rm AGN}\left(t\right)$. In most cases, when the outflow
properties change gradually, the results of the numerical integration
do not depend on the chosen integration scheme; we checked that this
is so using a third order Taylor integration scheme and both DKD and
KDK leapfrog algorithms. These methods produce slight differences in
situations where one or more parameters change abruptly and rapidly,
but these situations occur only due to the assumed perfect
conservation of energy and would not appear in reality, so the choice
of the integrator should not introduce errors larger than our other
assumptions do. Therefore, the calculations presented below are
performed using a third order Taylor integration scheme for outflow
radius (and, correspondingly, a second order scheme for velocity and a
first order scheme for acceleration). The timestep is adaptive and
equal to $\Delta t = f_{\rm CFL} {\rm min}\lbrace R/\dot{R},
\dot{R}/\ddot{R}, \ddot{R}/\dddot{R}\rbrace$, where $f_{\rm CFL} =
0.1$ is the Courant factor.

The equation of motion has a large number of free parameters
specifying the gas and background density distributions and the AGN
luminosity, therefore we first analyze the behaviour of a
fiducial model and then consider deviations from it. The fiducial
model has a galaxy composed of an $M_{\rm h} = 6 \times 10^{11} \;
\msun$ halo which has an NFW \citep{Navarro1997ApJ} density profile
with virial radius $r_{\rm vir} = 200$~kpc and concentration parameter
$c = 10$, and an $M = 4 \times 10^{10} \;\msun$ bulge with a Hernquist
density profile with scale radius $r_{\rm b} = 1$~kpc. The gas
fraction in the halo is $f_{\rm g,h} = 10^{-3}$, while in the bulge it
is $f_{\rm g,b} = 0.8$. The central AGN is powered by accretion on to
a black hole of mass $M_{\rm BH} = 2 \times 10^8 \;\msun$.

We use the same initial conditions for the outflow properties in each
simulation, with $R_{\rm 0} = 0.01$~kpc, $\dot{R}_{\rm 0} = 400 \,{\rm
  km s}^{-1}$ and $\ddot{R}_{\rm 0} = 0$. In earlier work
\citep{King2011MNRAS}, we showed that the initial conditions have
little effect upon the simulation results, provided that the outflow
is initially confined to the central parts of the galaxy, as is done
here.

We consider a schematic AGN luminosity history describing three stages
of SMBH growth. In the first stage, the SMBH grows together with its
host galaxy, but generally stays below the mass given by the relevant
$M-\sigma$ relation. Some small--scale outflows might be formed during
this stage during periods of AGN activity, but these do not have a
large--scale impact upon the gas distribution in the
galaxy. Accordingly, we do not model this stage. The second stage
occurs when episodic AGN activity inflates a large--scale outflow. We
model this as a `flickering' AGN, with luminosity equal to $L_{\rm
  Edd}$ for periods of $t_{\rm on} = 5\times10^4$~yr, separated by
some inactive phase, giving an AGN duty cycle $f_{\rm AGN}$
\citep{King2008MNRAS, Schawinski2015MNRAS, King2015MNRAS}.  Using
different values of $t_{\rm on}$ has no effect on our overall
results. At the end of this stage, the AGN has removed enough gas from
its surroundings to completely shut off its own accretion, and so the
third stage is marked by a coasting outflow which is not illuminated
by an AGN. This history is qualitatively consistent with the
cosmological quasar number density function
\citep[e.g.,][]{Croom2004MNRAS, Richards2006AJ}, which shows quasar
numbers increasing from high redshifts to $z \simeq 2$ and decreasing
after this. In our picture, the increase in quasar numbers happens as
SMBHs grow to higher masses, while the decrease occurs once they begin
quenching their own gas supplies.

\begin{figure}
  \centering
    \includegraphics[trim = 0 0 0 0, clip, width=0.45 \textwidth]{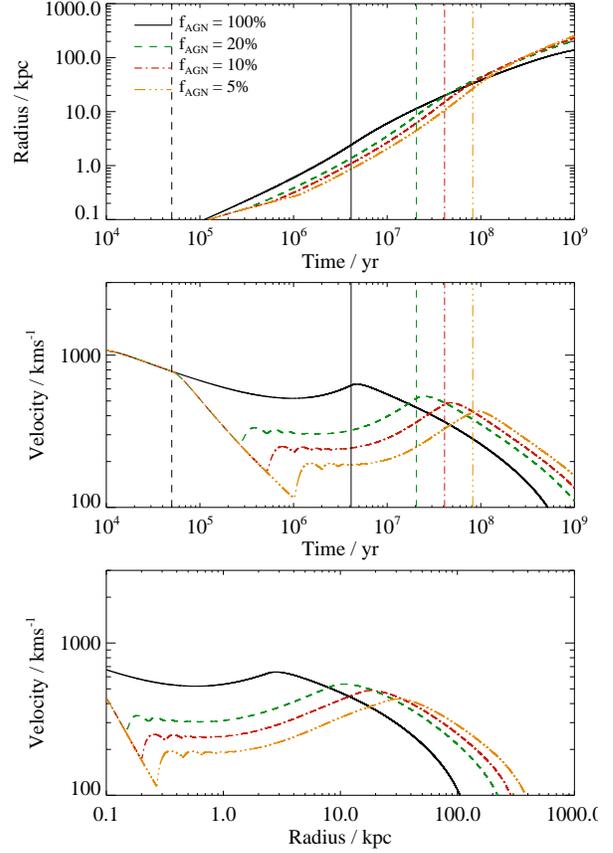}
  \caption{Outflow radius against time (top), velocity against time
    (middle) and velocity against radius (bottom) plots for the
    fiducial galaxy parameters (see text) and AGN activity
    characterised by $t_{\rm on} = 5\times10^4$~yr bursts followed by
    periods of inactivity giving duty cycles of $100\%$, $20\%$,
    $10\%$ and $5\%$ for black solid, green dashed, red dot-dashed and
    orange triple-dot-dashed lines, respectively. The thin
      vertical dashed line indicates the end of the first AGN activity
      episode and the four thicker vertical lines indicate the final
      end of AGN activity in the corresponding models. The AGN is
    switched off completely after releasing an amount of energy equal
    to $1.5/\left(\eta/2\right)$ times the binding energy of the gas.}
  \label{fig:fiducial_velocities}
\end{figure}

\begin{figure}
  \centering
    \includegraphics[trim = 0 0 0 0, clip, width=0.45 \textwidth]{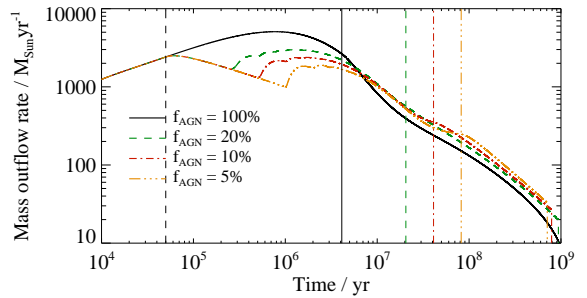}
  \caption{Mass outflow rates for the four models with fiducial galaxy
    parameters (same as in Figure
    \ref{fig:fiducial_velocities}). Colours and line styles are the
    same as in Figure \ref{fig:fiducial_velocities}.}
  \label{fig:fiducial_massrates}
\end{figure}

The precise value of AGN energy injection required to quench accretion
is difficult to determine. Since the AGN may well be fed by dense gas
clumps which are relatively unperturbed by the passage of an
energy--driven outflow \citep{Zubovas2014MNRASa}, injection greater
than the binding energy of a spherically--symmetric mass distribution
may be needed. In addition, the wind has not only to lift the gas
against gravity, but also expand itself, doing nonzero $P$d$V$ work in
the process. After some trial and error, we assumed a total luminous
energy release of $30$ times the gas binding energy in our fiducial
model.  This gives the wind kinetic energy driving the outflow as
$\eta/2 \times 30 = 1.5$ times the gravitational binding energy. For
the fiducial model, $E_{\rm b} = 1.2 \times 10^{59}$~erg, so the AGN
has to be active for $\sim 4.4$~Myr in total, assuming a constant
Eddington ratio $l = 1$ during the active phases. We also consider
total energy inputs 2 times higher and 2 times smaller than the
fiducial one. For each of these, we calculate the expansion of the
outflow with AGN duty cycles of $100\%$, $20\%$, $10\%$ and $5\%$.

\subsection{Fiducial model results}

\begin{figure}
  \centering
    \includegraphics[trim = 0 0 0 0, clip, width=0.45 \textwidth]{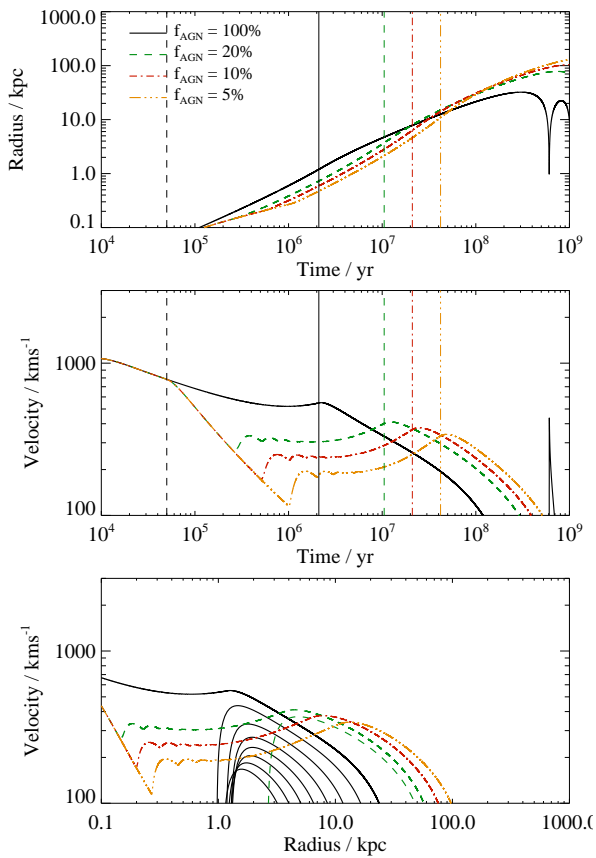}
  \caption{Same as Figure \ref{fig:fiducial_velocities}, but for total
    AGN energy injection equal to $0.75/\left(\eta/2\right)$ times the
    binding energy. The outflow stalls and oscillates, leading to
      the multiple black and green-dashed lines seen in the bottom
      panel.}
  \label{fig:lowen_velocities}
\end{figure}

\begin{figure}
  \centering
    \includegraphics[trim = 0 0 0 0, clip, width=0.45 \textwidth]{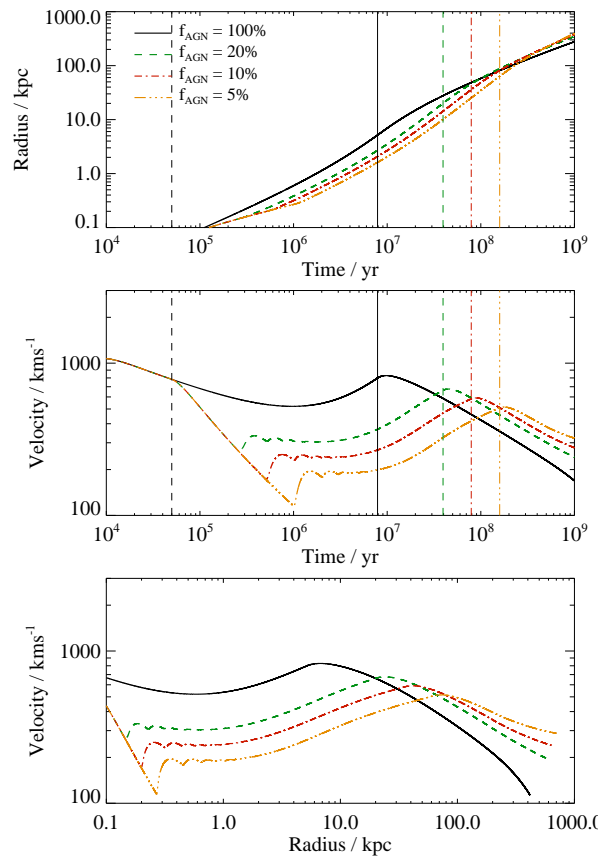}
  \caption{Same as Figure \ref{fig:fiducial_velocities}, but for total
    AGN energy injection equal to $3/\left(\eta/2\right)$ times the
    binding energy.}
  \label{fig:highen_velocities}
\end{figure}

In Figure \ref{fig:fiducial_velocities}, we show the time evolution of
the outflow radius (assumed to be given by the contact discontinuity
radius $R(t)$; top panel) and velocity $\dot R(t)$ (middle panel), and
the relation between velocity and radius (bottom panel) for the
fiducial model. The vertical dashed line shows the duration of a
single AGN activity episode. The AGN shuts off completely at $t =
4.1$, $20.5$, $41.0$ and $82.0$~Myr for the four AGN duty cycles
(black solid, green dashed, red dot--dashed and orange
dot--double--dashed lines, respectively). Qualitatively, all four
outflows expand in a similar fashion. The oscillations caused by AGN
flickering smooth out quickly (in $<3$\,Myr), and the outflow behaves
as if driven by an AGN with a lower but constant luminosity. The
outflow velocity does not vary significantly with time in any given
model and is typically between $200$ and $1000\,{\rm km\,s^{-1}}$ in
all the models considered. These velocities are slightly lower than
predicted in our previous work \citep{Zubovas2012ApJ} where we
considered outflows in an isothermal potential, and more in line with
observed AGN outflows (see Discussion).

Importantly, as shown by \citet{King2011MNRAS}, the outflows continue
to expand significantly after the AGN switches off.  When the AGN is
active, the outflow is within $r = 10$~kpc for $100\%$, $100\%$,
$72\%$ and $46\%$ of the full cycle, for duty cycles of $100\%$,
$20\%$, $10\%$ and $5\%$ respectively.  We suggest that this may
explain the typical spatial extent of manifestly AGN--driven outflows
of several kpc \citep[e.g.,][]{Shih2010ApJ} and the lack of such
outflows observed beyond $\sim 10$~kpc
\citep[e.g.][]{Spence2016MNRAS}. We return to this point in the
Discussion.

The mass outflow rate, defined as the time derivative of the baryonic
mass contained within $R$, is also large both while the AGN is active,
and for some time after it switches off. The values reached, $\dot{M}
> 10^3 \msun$~yr$^{-1}$, should be seen as upper limits, because in
reality a significant fraction of the gas is dense and resists being
pushed away. Nevertheless, large mass outflow rates need not coincide
with AGN activity, therefore the lack of an active nucleus in a given
galaxy should not be taken as evidence that any observed outflow is
not AGN-driven.

Figures \ref{fig:lowen_velocities} and \ref{fig:highen_velocities},
show the same results as in Figure \ref{fig:fiducial_velocities}, but
for different values of limiting AGN input energy -- $0.75$ and $3$
times the binding energy respectively. We see that if the AGN is very
efficient at quenching its own mass supply (Figure
\ref{fig:lowen_velocities}), the outflows are unable to escape from
the halo completely and stall at a few tens to $100$~kpc, oscillating
afterwards (see the bottom panel). The AGN switches off before any
outflow escapes beyond $10$~kpc, so we do not expect to see outflows
with an active AGN at radii $\gtrsim 10$\,kpc.  Conversely, a
long--lasting AGN episode (Figure \ref{fig:highen_velocities}) clears
gas out more easily.  The outflow maintains a velocity $\sim 200\,{\rm
  km\,s^{-1}}$ even at $R = 1$\,Mpc. In this case, the fraction of
outflows within $R = 10$\,kpc while the AGN is active is $100\%$,
$57\%$, $36\%$ and $24\%$ for the four duty cycles. Although some of
these outflows do get outside $R = 10$\,kpc with the AGN still active,
the fractions are still significantly lower than the $\sim 99\%$
expected for an outflow moving at a constant velocity out to $R =
1$\,Mpc.

\subsection{Trends with halo mass}

\begin{figure}
  \centering
    \includegraphics[trim = 0 0 0 0, clip, width=0.45 \textwidth]{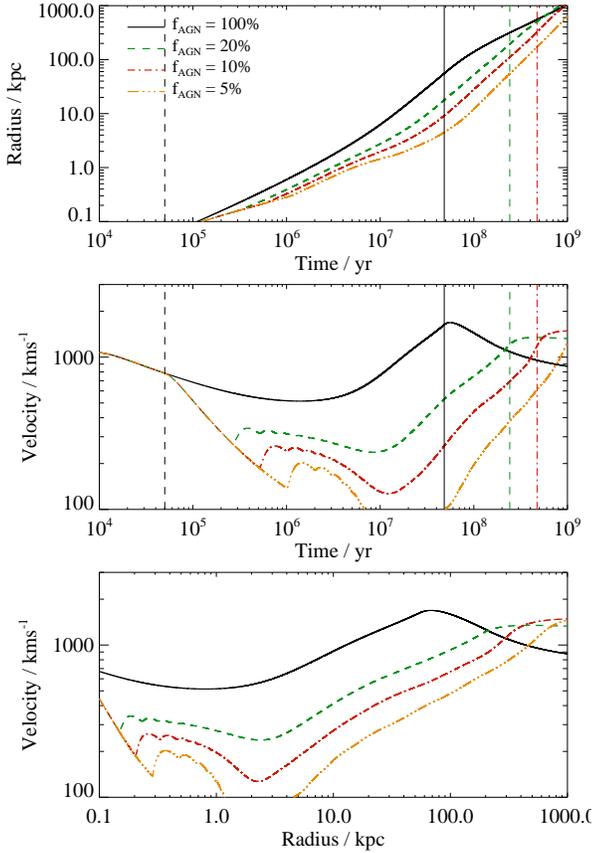}
  \caption{Same as Figure \ref{fig:fiducial_velocities}, but for a
    halo with mass $M_{\rm h} = 10^{13} \msun$.}
  \label{fig:highmass_velocities}
\end{figure}

\begin{figure}
  \centering
    \includegraphics[trim = 0 0 0 0, clip, width=0.45 \textwidth]{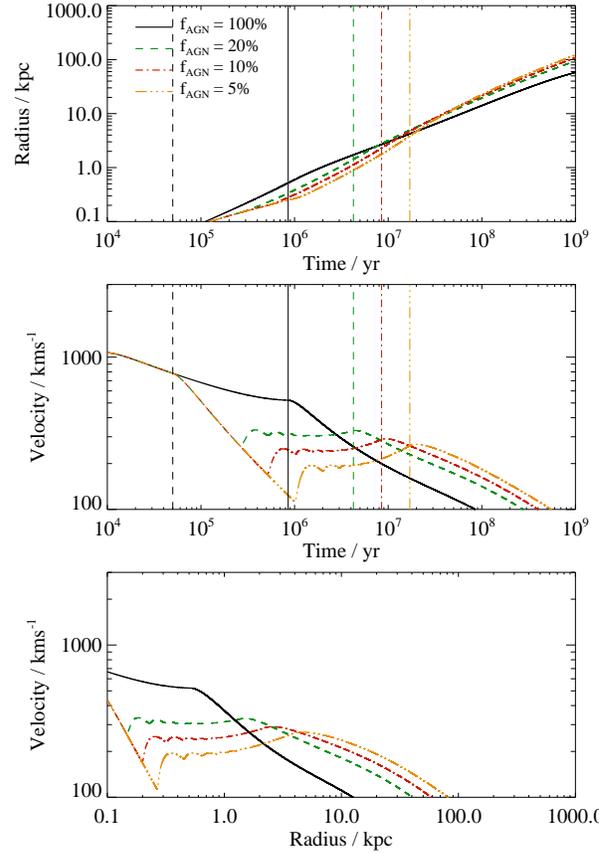}
  \caption{Same as Figure \ref{fig:fiducial_velocities}, but for a
    halo with mass $M_{\rm h} = 10^{11} \msun$.}
  \label{fig:lowmass_velocities}
\end{figure}

We also tested how outflows behave in galaxies with different masses.
Figures \ref{fig:highmass_velocities} and \ref{fig:lowmass_velocities}
show the results for galaxies with halo mass $10^{13} \msun$ and
$10^{11} \msun$.  We keep the same ratios of $M_{\rm bulge}/M_{\rm
  total}$ and $M_{\rm BH}/M_{\rm total}$ as well as gas fractions.

The outflows behave very similarly at early times for all halo
masses. This happens because in the central parts of the galaxy (small
$R$) the kinetic terms in the equation of motion (eq. \ref{eq:eom})
dominate the gravity terms, so the outflow expansion depends mainly on
the ratio $L_{\rm AGN}/M_{\rm g} \propto M_{\rm BH}/f_{\rm g}/M_{\rm
  total} = {\rm const}$. But the binding energy goes as $M^2$, so the
time for which the AGN is active is $\sim 10$ times higher for the
$M_{\rm h} = 10^{13} \msun$ halo. In the $M_{\rm h} = 10^{11} \msun$
simulation, we allow the AGN to remain active for long enough to
inject $3$ times the gravitational binding energy of the gas into the
outflow to ensure gas escape, so the activity is $\sim 5$ times
shorter than in the fiducial model.

In low--mass halo models, the outflow has only expanded to a few kpc by
the time the AGN switches off, and later coasts with a gradually
decreasing velocity. Outside $\sim 10$\,kpc the outflow velocity drops
below $100\,{\rm km\,s^{-1}}$, making it very difficult to
detect. 

Conversely, in models with high--mass halos the fraction of outflows
within $R < 10$~kpc during phases of AGN activity is $44\%$, $20\%$,
$16\%$ and $14\%$ for the four AGN duty cycles, respectively. The
outflow velocity reaches $\sim 1000\,{\rm km\,s^{-1}}$ before the AGN
switches off for the final time, so in this case, outflowing gas would
be detectable out to large distances.

\section{Induced starbursts in galaxy discs}\label{sec:burst}

\begin{figure}
  \centering
    \includegraphics[trim = 0 0 0 0, clip, width=0.45 \textwidth]{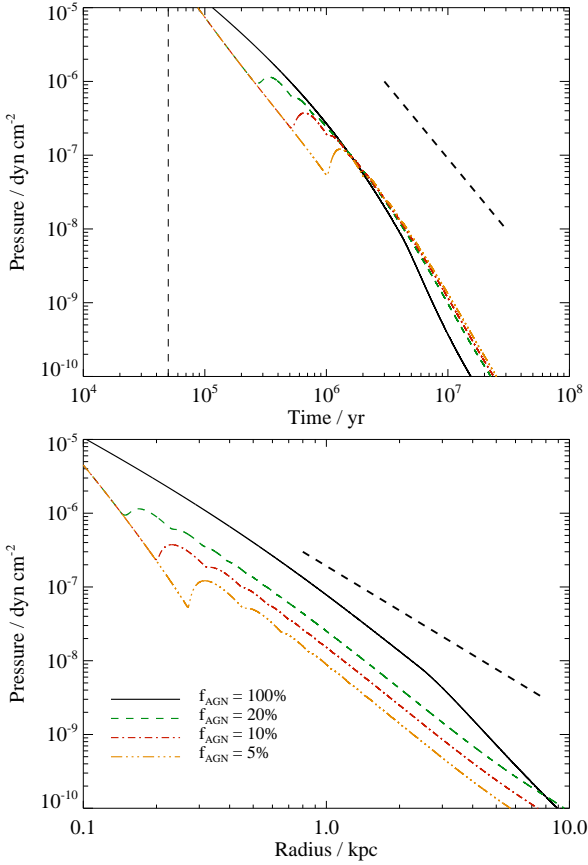}
  \caption{Outflow forward shock pressure against time (top) and
    radius (bottom). Lines indicate models with AGN duty cycles of
    $100\%$ (black solid), $20\%$ (green dashed), $10\%$ (red
    dot-dashed) and $5\%$ (yellow triple-dot-dashed). The AGN is switched
    off completely after releasing $1.5/\left(\eta/2\right)$ times the
    binding energy of the gas. Thick dashed lines show $p \propto
    t^{-2}$ and $p \propto R^{-2}$ dependencies.}
  \label{fig:fiducial_pressures}
\end{figure}

So far we have discussed the expansion of an outflow driven by a
flickering, AGN whose total energy output is limited. We now consider
the effects of the outflow pressure on the galaxy disc. We set up a
disc extending from $0.125$ to $40$~kpc in radius, with an
exponentially decreasing radial density profile \citep[consistent with
  observations, e.g.,][]{Bigiel2012ApJ, Wang2014MNRAS}, with a scale
length of $3$~kpc and a total mass $M_{\rm d} = 3.6\times10^{11}
\msun$. We consider several values of the disc gas fraction $f_{\rm
  g,d}$: $100\%$, for a system at high redshift; $50\%$, for a galaxy
which has formed roughly half of its stars (fiducial model); and
$20\%$ and $10\%$, representing galaxies in the Local Universe. We
calculate the star formation rate based on the volumetric
Kennicutt--Schmidt relation:
\begin{equation}
\rho_{\rm SFR} = \epsilon_* \frac{\rho_{\rm g}}{t_{\rm ff}} =
\epsilon_* \left(\frac{32G}{3\pi}\right)^{1/2} \rho_{\rm g}^{3/2},
\end{equation}
which can be recast for simplicity into a relation based on surface
densities:
\begin{equation} \label{eq:sigma_sfr_sp}
  \begin{split}
    \Sigma_{\rm SFR, sp} &= \epsilon_*
    \left(\frac{32G}{3\pi}\right)^{1/2} \Sigma_{\rm g}^{3/2} h_{\rm
      d}^{-1/2} \\ & \simeq 4.5\times 10^{-3} \Sigma_{10}^{3/2}
    h_{300}^{-1/2} \epsilon_{0.02} \msun {\rm yr}^{-1} {\rm kpc}^{-2},
  \end{split}
\end{equation}
where $\epsilon \equiv 0.02 \epsilon_{0.02}$ is the star formation
efficiency per dynamical time, $\rho_{\rm g}$ is the gas volume
density, $\Sigma_{\rm g} \equiv 10 \Sigma_{10} \msun {\rm pc}^{-2}$ is
the gas surface density and $h_{\rm d} \equiv 300 h_{300}$~pc is the
scale height of the disc. We choose a value for the scale height
which gives the appropriate scaling for the Kennicutt--Schmidt
relation \citep{Kennicutt1998ApJ}.

External compression caused by the passing outflow can stimulate star
formation in the disc by reducing the effective dynamical time of
star-forming clouds \citep{Zubovas2014MNRASc}. We parameterize this
as 
\begin{equation} \label{eq:ind_sfr}
\Sigma_{\rm SFR, ind} = \Sigma_{\rm SFR, sp}\left[\left(1+\frac{P_{\rm
    ext}}{P_{\rm d}}\right)^{\rm \beta}-1\right],
\end{equation}
where
\begin{equation}
P_{\rm d} = \rho_{\rm g} c_{\rm s}^2 = \Sigma_{\rm g} h_{\rm d}
\left(\frac{v_{\phi}\left(r\right)}{r}\right)^2
\end{equation}
is the disc pressure at radius $r$, and $\beta$ is a free parameter of
the model. $P_{\rm ext}$ is the external pressure acting upon the disc
(see below), which may be higher or lower than the disc pressure, but
would act as an additional compressive term either way. In the
simplest case, where external pressure is added linearly to the
gravitational energy density, $\beta = 1$, but effects such as
pressure instabilities following passage of a shock wave
\citep{Hopkins2010MNRAS} and/or the formation of instabilities behind
the shockwave \citep{Zubovas2014MNRASc} may decrease ($\beta <1$) or
increase ($\beta >1$) the SFR enhancement respectively.

To calculate the external pressure we consider the outflow to be
composed of two regions: inside the contact discontinuity, the
pressure is constant and equal to $P_{\rm cd}$, while outside the
discontinuity it changes linearly between $P_{\rm cd}$ and $P_{\rm
  out}$, where the two pressures are calculated as in \citep[cf.][ and
  equation \ref{eq:pres_cd}]{Zubovas2013MNRAS}:
\begin{equation}
P_{\rm cd} = \frac{1}{4\pi R^2}\left[\frac{{\rm d}}{{\rm
      d}t}\left[M\dot{R}\right]+\frac{GM\left(M_{\rm b} +
    M/2\right)}{R^2}\right],
\end{equation}
\begin{equation}
P_{\rm out} = \frac{4}{3} \rho_{\rm amb} \dot{R}^2.
\end{equation}
Here the masses include only material within $R$, and $\rho_{\rm amb}$
is the density of the ambient medium at radius $R$. In writing the
expression of $P_{\rm out}$, we make an assumption that the radius of
the outer shock is $4/3$ times the radius of the contact
discontinuity. This is a reasonable assumption so long as the velocity
of the outflow is significantly higher than the sound speed in the gas
\citep[cf.][]{Zubovas2012ApJ}. We plot the evolution of $P_{\rm out}$
with time and radius in the top and bottom panels of Figure
\ref{fig:fiducial_pressures}, respectively. It is worth noting that
the radiation pressure $p_{\rm rad} = L/\left(4 \pi R^2 c\right)$ is
$\sim 15$ times lower than $P_{\rm out}$ even when $L = L_{\rm Edd}$,
i.e. the outflow has a considerably higher pressure than the radiation
field driving it.

\begin{figure}
  \centering
    \includegraphics[trim = 0 0 0 0, clip, width=0.45 \textwidth]{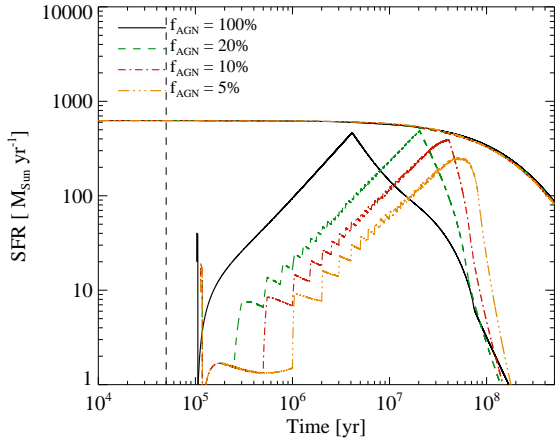}
  \caption{Star formation rate versus time for the models with varying
    AGN duty cycles: $100\%$ (black solid), $20\%$ (green dashed),
    $10\%$ (red dot-dashed) and $5\%$ (yellow triple-dot-dashed). The
    gas fraction in the disc is $50\%$ and the total disc mass is $3.6
    \times 10^{11} \msun$. The AGN is switched off completely after
    releasing $1.5/\left(\eta/2\right)$ times the binding energy of
    the gas. Thin lines represent spontaneous star formation in the
    corresponding models.}
  \label{fig:fiducial_sfr}
\end{figure}

\begin{figure}
  \centering
    \includegraphics[trim = 0 0 0 0, clip, width=0.45 \textwidth]{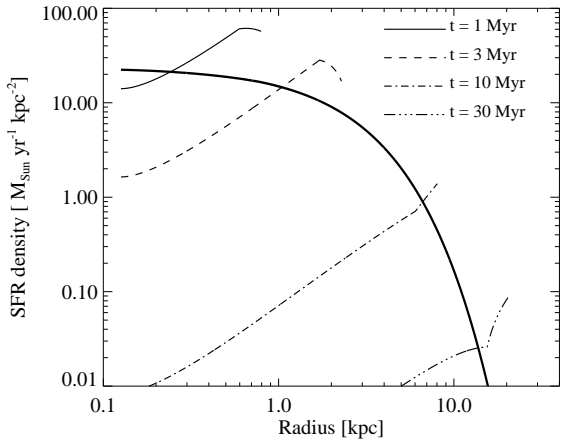}
    \caption{SFR density as function of radius in the fiducial model
      with $100\%$ AGN duty cycle. Solid line: $t = 1$~Myr after the
      AGN switches on; dashed line: $t = 3$~Myr; dot-dashed line: $t =
      10$~Myr; triple-dot-dashed line: $t = 100$~Myr. Thick solid line
      shows the spontaneous SFR density profile at $t = 1$~Myr; this
      profile remains qualitatively similar, but globally decreases,
      at later times.}
  \label{fig:fiducial_sfr_radial}
\end{figure}

\begin{figure}
  \centering
    \includegraphics[trim = 0 0 0 0, clip, width=0.45 \textwidth]{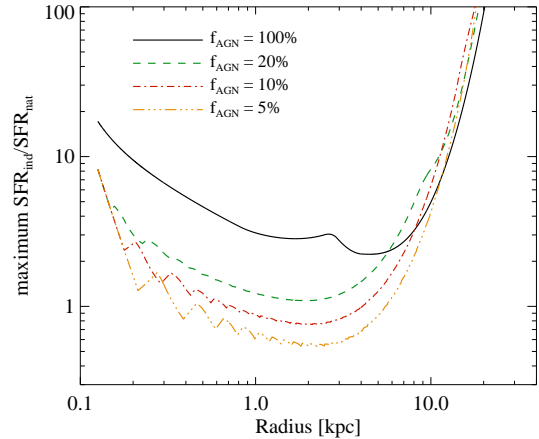}
    \caption{Maximum ratio of induced to spontaneous SFR density as
      function of radius in the fiducial models. Line styles and
      colours as in Figure \ref{fig:fiducial_sfr}.}
  \label{fig:fiducial_burstmax}
\end{figure}

\subsection{Fiducial model results}

We give first the results of the fiducial model (disc gas fraction
$50\%$) with four AGN duty cycles, as above. Figure
\ref{fig:fiducial_sfr} shows the rates of induced star formation
(thick lines) as functions of time. The thin lines represent the
`spontaneous' star formation rate $\dot{M}_{\rm SFR,sp}$, i.e. that
given by integrating the KS law (eq. \ref{eq:sigma_sfr_sp}) over the
whole disc; this SFR decreases with time as gas is consumed in the
disc. In the model with continuous AGN activity (black solid line),
induced star formation keeps growing until the AGN switches off at $t
\simeq 4.1$~Myr, but never reaches the level of spontaneous star
formation, falling short by about a factor of 2. The situation is
similar for the lower AGN duty cycles, except that there the induced
SFR fluctuates significantly between the `on' and `off' phases of the
AGN. This suggests that AGN--induced star formation would be difficult
to detect in an unresolved galaxy where only the integrated SFR could
be estimated.

If the galaxy disc can be resolved, AGN effects upon star formation
are more easily detected. Figure \ref{fig:fiducial_sfr_radial} shows
the radial SFR density profiles of spontaneous (thick solid line) and
induced (thin lines) star formation at different times from 1 to
30~Myr. As expected, the induced SFR at a given radius decreases with
time as the outflow expands and the pressures at the centre and the
contact discontinuity drop. The induced SFR increases with radius out
to the contact discontinuity at all times. This happens because even
though the outflow pressure is uniform inside the contact
discontinuity, the internal disc pressure decreases with decreasing
$\Sigma_{\rm g}$ and radius, so the outflow pressure becomes more
significant. The outflow-to-disc pressure ratio increases from
typically negligible values close to the centre to $P_{\rm ext}/P_{\rm
  d} \sim 3$ at the outer edge of the outflow. Conversely, as long as
the AGN is active, the pressure at the outer shock decays with time
more slowly than the pressure at the contact discontinuity because of
the increase in outflow velocity, so the ratio of induced SFR density
at the outermost point in the disc affected by the outflow to the
induced SFR density at the contact discontinuity increases with
time. Together, these effects lead to a situation where the star
formation enhancement is more significant in the outskirts of the disc
than the centre.

We show this result in Figure \ref{fig:fiducial_burstmax}, where we
plot the maximum ratio of the value $\Sigma_{\rm SFR,ind}/\Sigma_{\rm
  SFR, nat}$ at every radius in the disc attained before the AGN
finally switches off. In the central kiloparsec, this ratio decreases
with increasing radius and is only of order a few, but beyond $r
\simeq 2-6$~kpc, depending on $f_{\rm AGN}$, it starts growing and
reaches a value $> 100$ at $\sim 20$~kpc, growing slightly faster
with decreasing AGN duty cycle. The fiducial models reveal that
AGN outflows can significantly increase the galactic disc SFR within
the central kiloparsec and outside $\sim 10$~kpc, but the effect on
integrated SFR parameters is only moderate.

\subsection{Star formation triggering in discs with different gas content}

\begin{figure}
  \centering
    \includegraphics[trim = 0 0 0 0, clip, width=0.45 \textwidth]{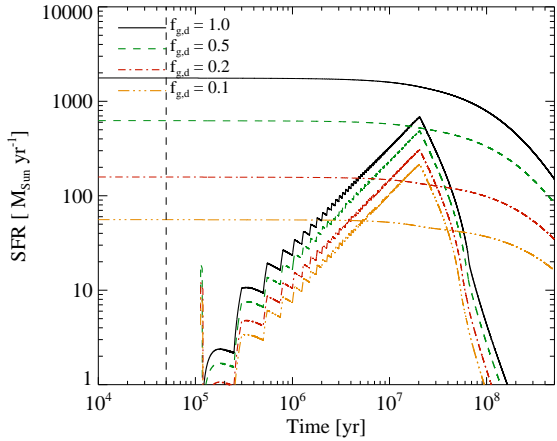}
  \caption{Star formation rate versus time for the models with varying
    disc gas fractions: $100\%$ (black solid), $50\%$ (green dashed),
    $20\%$ (red dot-dashed) and $10\%$ (yellow triple-dot-dashed). The
    AGN is switched off completely after releasing
    $1.5/\left(\eta/2\right)$ times the binding energy of the
    gas. Thin lines represent spontaneous star formation in the
    corresponding models.}
  \label{fig:lowgas_sfr}
\end{figure}

\begin{figure}
  \centering
    \includegraphics[trim = 0 0 0 0, clip, width=0.45 \textwidth]{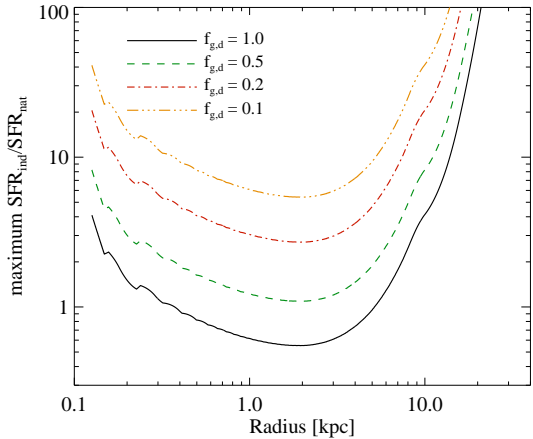}
  \caption{Maximum ratio of induced to spontaneous SFR density as function
    of radius in the models with varying disc gas fraction. Line
    styles and colours as in Figure \ref{fig:lowgas_sfr}.}
  \label{fig:lowgas_burstmax}
\end{figure}

We next consider the effects of an AGN outflow upon galaxy discs with
different gas fractions. We present results only for a model with
$f_{\rm AGN} = 20\%$, as we believe this to be the most likely of the
four AGN duty cycles we investigate (see Section \ref{sec:detect},
below), and models with other values of $f_{\rm AGN}$ show
qualitatively similar behaviour (see Figure
\ref{fig:fiducial_sfr}). The expected trend with changing gas density
is that the spontaneous SFR decreases as $\Sigma_{\rm g}^{3/2}$, while
the induced SFR decreases only as $\Sigma_{\rm SFR, nat}/P_{\rm d}
\propto \Sigma_{\rm g}^{1/2}$. The ratio of the two increases with
decreasing gas fraction as $\Sigma_{\rm g} \propto f_{\rm g,\ d}$.

Figure \ref{fig:lowgas_sfr} shows the star formation histories of
models with four different gas fractions. As before, thin lines are
spontaneous star formation, while thick ones show induced SFR. As
expected, the spontaneous star formation rate declines as $f_{\rm
  g,\ d}^{1.5}$. The induced star formation rate declines slightly
faster than $f_{\rm g, \ d}^{0.5}$, mainly because gas is completely
consumed in some annuli of the disc in the lower density simulations,
further reducing the integrated SFR. For gas fractions $0.2$ and
$0.1$, corresponding to initial disc gas masses of $7.2$ and $3.6
\times 10^{10} \msun$, the induced SFR is higher than the spontaneous
SFR for several Myr. Importantly, the SFR enhancement persists after
the AGN switches off, as the outflow is still moving through the
galaxy. So in gas--poor galaxy discs AGN activity may induce
starbursts which might be visible even in integrated SFR data as
offsets from the KS relation.

The ratio of induced to spontaneous SFR is presented in Figure
\ref{fig:lowgas_burstmax}. The expected trend is visible there as
well, with induced starbursts in low--density galaxy discs increasing
the star formation efficiency by more than an order of magnitude from
the centre to the outskirts. All four models show qualitatively
identical behaviour. The ratio decreases out to $\sim 5$~kpc and then
increases further out. The bump at $R \simeq 2.5$~kpc appears because
the AGN switches off when the outflow has expanded to that distance,
and so the outflow pressure begins to drop rapidly.

\subsection{Alternative prescriptions for induced star formation}

\begin{figure}
  \centering
    \includegraphics[trim = 0 0 0 0, clip, width=0.45 \textwidth]{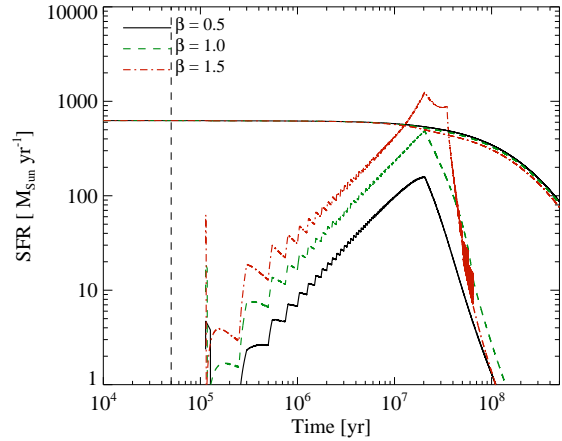}
  \caption{Star formation rate versus time for the models with $50\%$
    disc gas fraction and varying star formation induction parameter
    $\beta$: $\beta=0.5$ (black solid), $\beta=1$ (green dashed) and
    $\beta=1.5$ (red dot-dashed). The AGN is switched off completely
    after releasing $1.5/\left(\eta/2\right)$ times the binding energy
    of the gas. Thin lines represent spontaneous star formation in the
    corresponding models.}
  \label{fig:varbeta_sfr}
\end{figure}

\begin{figure}
  \centering
    \includegraphics[trim = 0 0 0 0, clip, width=0.45 \textwidth]{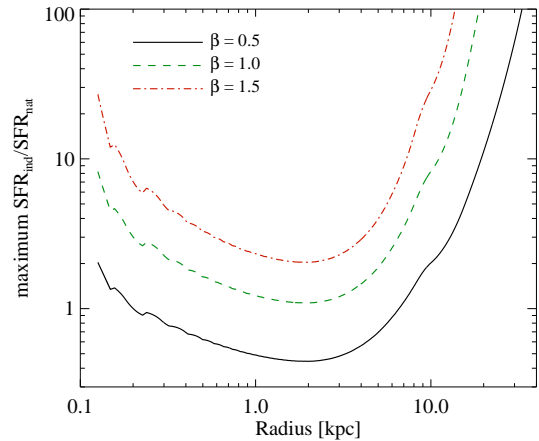}
  \caption{Maximum ratio of induced to spontaneous SFR density as function
    of radius in the models with varying $\beta$. Line styles and
    colours as in Figure \ref{fig:varbeta_sfr}.}
  \label{fig:varbeta_burstmax}
\end{figure}

Given that the precise effect of the outflow compression upon the ISM
is uncertain, we test four different power--law indices $\beta$ for the
calculation of the induced SFR (eq. \ref{eq:ind_sfr}). The results for
$\beta = 0.5$, $1$ and $1.5$ are shown in Figure
\ref{fig:varbeta_sfr}. As expected, the induced SFR is higher for
higher values of $\beta$. If $\beta = 0.5$, i.e. the disc resists
compression via processes other than star formation feedback, the
induced SFR is $\sim 2-3$ times lower than in the fiducial model. On
the other hand, $\beta = 1.5$ produces induced SFR becoming higher
than the spontaneous SFR at the peak at $t \simeq 4$~Myr. This induced SFR
value is a factor $\sim 2$ higher than in the fiducial model for the
first $\sim 10$~Myr. Later, as the outflow expands to the disc
outskirts, SFR induction is so efficient that the total induced
SFR begins to rise with time. The second peak at $t \sim 40$~Myr
represents the time when gas is completely consumed in some disc
annuli around $R = 20-22$~kpc. As star formation stops there, the
total induced SFR drops significantly as well.

The radial induced starburst plot (Figure \ref{fig:varbeta_burstmax})
shows the same qualitative behaviour for all three models, with the
weakest starburst induced just outside the bulge ($R = 2-7$~kpc) and
increasing both inward and outward. Even a reduced--efficiency ($\beta
= 0.5$) SFR induction can result in a noticeable starburst at $r >
20$~kpc -- the derived star formation efficiency is an order of
magnitude greater than predicted by the KS law. The extreme ratios
achieved for the enhanced--efficiency ($\beta = 1.5$) induction model
provoke rapid consumption of gas in the outskirts of the disc, and
such powerful starbursts may be difficult to detect because of their short
lifetimes.

\section{Discussion}\label{sec:discuss}

\subsection{What is the likely duty cycle of AGN? Outflow detectability} \label{sec:detect}

By considering the fraction of AGN host galaxies which have not yet
been ionised by the X--rays and UV from the nucleus,
\citet{Schawinski2015MNRAS} infer that AGN phases typically last only
$\sim 10^5$~yr. A similar result is reached by considering the time
required to accrete a disc with outer edge (and hence mass) bounded by
self-gravity \citep{King2015MNRAS}. In our model, a single short phase
like this is certainly not enough to inflate a massive outflow, so
multiple phases of AGN activity are required. Our results depend on
the fraction of time that the AGN is active, and we consider simple
arguments constraining the possible values of this fraction.
 
Given the observed fraction of galaxies which are active
\citep[e.g.,][]{Xue2010ApJ} we should have $f_{\rm AGN} \sim 5\%$
averaged over the age of the galaxy. A similar fraction comes from the
Soltan argument \citep{Soltan1982MNRAS}. If an SMBH grows from stellar
$M_{\rm init} \sim 10 \msun$ to its present mass, it must spend at
least ${\rm ln}\left(M_{\rm fin}/M_{\rm init}\right) \sim 16$ Salpeter
times accreting material. This duration is $16 t_{\rm Sal} \sim
670$~Myr$\sim 0.05t_{\rm H}$, where $t_{\rm H}$ is the Hubble
time. However, as the galaxy evolves, the frequency of AGN activity
evolves as well. At high redshift there is a lot of gas available to
feed the SMBH and the galaxy may be active more often, while at low
redshift it is hardly ever active. The phase of gas ejection via a
massive outflow marks the end of the high--activity phase of galaxy
evolution, and so should be marked by a typical $f_{\rm AGN}$ larger
than the long-term average. So our simulations with $f_{\rm AGN} =
5\%$ represent a lower limit to the expected AGN fraction at this
stage of galaxy evolution. We expect $f_{\rm AGN} \sim10\%, 20\% $ to
be more representative of reality, but the actual value is uncertain.

Our results then suggest that massive outflows are typically detected
close to AGN because AGN switch off for good by the time outflows move
far out from the nucleus. Outflow signatures might be detectable
around inactive galaxies at higher distances than around active
galaxies, however as the outflow expands without driving, its density
and velocity decrease and it becomes progressively more difficult to
detect. Outflows might also be detectable further from the AGN in more
massive galaxies than in lower--mass ones, where the AGN needs to be
active for longer and drive the outflow further away to ensure
quenching of the mass supply.

Another interesting consequence of having $f_{\rm AGN} < 1$ is that
even in the central parts of galaxies, AGN-driven outflows might often
be detected without the presence of the driving AGN itself. These
outflows might be erroneously identified as driven by star-formation
activity. Therefore, we predict that a larger fraction of outflows
happen due to AGN activity than expected from a naive interpretation
of observations.

It is important to consider how these results depend on the
assumptions we made regarding the model. Real outflows are not uniform
nor spherically symmetric, therefore there is a spread of radii at
which the outflow is observed at any given time. The spread is
mediated by Rayleigh-Taylor instabilities \citep{King2010MNRASb} and
by the presence of dense material in the undisturbed ISM, thus the
outflowing gas should be seen both closer to the AGN and further away
from it than the formal radius of the contact discontinuity. The dense
molecular gas, which is the usual tracer of these outflows, is more
likely to lag behind the contact discontinuity than ahead of it
\citep[see, e.g., the numerical simulations
  by][]{Nayakshin2012MNRASb}, therefore our conclusion that outflows
should be detected close to their parent AGN is only strengthened.

Another issue is the total energy input by the AGN required to shut
off its own accretion. In principle, the dense gas in the galaxy might
not be removed even if the outflow clears the diffuse gas, and thus
AGN might still be fed by infalling clouds. However, several processes
might prevent long-term feeding of an AGN after a large-scale outflow
is driven away:

\begin{itemize}

\item The fraction of dense gas in galaxy bulges is not very large to
  begin with, $M_{\rm dense}/M_{\rm gas} \lesssim 0.5$
  \citep{Blanton2009ARA&A, Saintonge2011MNRAS,
    Popping2014MNRAS}. Therefore, the outflow removes more than half of
  the material that might eventually feed the AGN.

\item The outflow moves past molecular clouds and either disperses
  them \citep{Hopkins2010MNRAS} or compresses them and enhances star
  formation \citep{Zubovas2014MNRASc}, further depleting the amount of
  dense gas available to feed the AGN.
  
\item Once the diffuse gas is gone, molecular gas is not replenished
  any more.

\end{itemize}

Taken together, these points suggest that most of the AGN feeding
reservoir is depleted, or at least significantly diminished, by the
large-scale outflow. In addition, our choice for basing the total AGN
energy release on the binding energy of the gas is consistent with
numerical hydrodynamical simulations which suggest that dark matter
halo masses, and hence the binding energies of those haloes, are the
most important parameter setting the mass of the SMBH
\citep{Booth2010MNRAS}. Therefore we are confident in suggesting that
once the AGN outflow unbinds most of the gas from the galaxy, the AGN
switches off, essentially, forever.

\subsection{Induced starbursts in nuclear gas rings}

Starbursts induced by the AGN outflow are stronger in the central
kiloparsec compared with the intermediate regions (Figures
\ref{fig:fiducial_burstmax}, \ref{fig:lowgas_burstmax} and
\ref{fig:varbeta_burstmax}). This happens because the outflow pressure is
very high there, while the disc gas density, and hence its pressure,
does not vary strongly within the central scale length.

These sub--kpc scales fall within the bulge of the galaxy, and
correspond to locations of nuclear gas and stellar rings seen in many
barred galaxies \citep[e.g.,][]{Kormendy2004ARA&A, Knapen2005A&A,
  Boeker2008AJ}. Gas in the rings may be directly affected by the
radial push of the AGN wind \citep{Zubovas2015MNRAS}, but could also
be compressed vertically and produce strong bursts of star formation,
as observed \citep{Allard2006MNRAS, Sarzi2007MNRAS}. However, any such
burst would probably be short, lasting for only a few Myr before the
nuclear ring relaxes to a baseline state, and so is probably
detectable only simultaneously with AGN activity or soon after the AGN
has switched off. Bulge stars and gas may also mask the effect of the
outflow on the nuclear gas ring, hampering simple interpretation of
the observations.

\subsection{Starbursts on galactic outskirts} \label{sec:outskirts}

In the outskirts of galaxy discs, gas density is low, resulting in low
pressure and low spontaneous star formation rate density. Therefore,
even though the induced star formation rate density decreases as well
(see Figure \ref{fig:fiducial_sfr_radial}), the ratio between the two
grows once the outflow moves beyond $\sim 5$~kpc (see Figure
\ref{fig:fiducial_burstmax}). Outside $R \simeq 10$~kpc, this ratio is
more than an order of magnitude. Such an offset from the typical
values predicted by the KS law should be easily detected, provided
that star formation in the galaxy is spatially resolved. Some examples
of such galaxies might be the ring galaxies such as the
Cartwheel. Detection of rings of star formation with very high star
formation efficiency in galactic outskirts might be taken as evidence
of recent (less than a few times $10^7$~yr) episode of nuclear
activity in the galaxy, especially since the star formation efficiency
generally tends to decrease in galactic outskirts \citep{Leroy2008AJ}.

\subsection{Inside-out quenching of star formation in galaxy discs}

Recently, \citet{Tacchella2015Sci} found evidence that star formation
in galaxy discs might be quenched from the inside out. This quenching
might simply be the result of faster gas depletion in the central
parts of the galaxy due to differences in dynamical time. On the other
hand, starbursts induced by AGN outflows might enhance this effect.

The induced star formation rate depends strongly on the outflow
pressure, which, in turn, depends on its velocity (see Figure
\ref{fig:fiducial_pressures}). As a result, when the outflow is driven
by a flickering AGN, induced star formation rate rises periodically
during the active phases, and decays during the inactive phases, even
though the outflow itself keeps expanding (see Figures
\ref{fig:fiducial_sfr}, \ref{fig:lowgas_sfr} and
\ref{fig:varbeta_sfr}). During each active phase, the outflow
encompasses a greater part of the disc and leads to faster consumption
of gas there. If the gas fraction in the disc is already low, the
induced star formation might consume gas and quench star formation in
the disc from the inside out. This process would manifest itself
afterward by a radial gradient of stellar ages, however, since the
outflow expands at a few hundred km/s, i.e. covers 1 kpc in a few Myr,
the gradient might be too shallow to detect easily.

\subsection{Flaring of discs in response to outflow passage}

One more effect that the outflow might have upon a galaxy gas disc is
causing it to flare in response to the outflow passage. This can be
seen by considering the following simple argument.

The pressure wave induced by the passing outflow moves vertically
through the disc with a velocity
\begin{equation}
  v_{\rm sh} \sim \sqrt{\frac{P_{\rm ext}}{P_{\rm d}}} c_{\rm s},
\end{equation}
and takes a time
\begin{equation}
  t_{\rm sh} \sim \frac{2H}{v_{\rm sh}} \sim 2 \sqrt{\frac{P_{\rm d}}{P_{\rm ext}}} t_{\rm dyn}
\end{equation}
to reflect back to the edge of the disc. In this time, the edge of the
outflow has moved outward a distance
\begin{equation}
  \Delta R_{\rm out} \sim v_{\rm out} t_{\rm sh} \sim 2\frac{v_{\rm out}}{\sigma} \sqrt{\frac{P_{\rm d}}{P_{\rm ext}}} \sigma t_{\rm dyn} \sim 2\frac{v_{\rm out}}{\sigma} \sqrt{\frac{P_{\rm d}}{P_{\rm ext}}} R,
\end{equation}
where $R$ is the original outflow location. Putting in typical values
$v_{\rm out}/\sigma \sim 5$, $P_{\rm ext}/P_{\rm d} \sim 100$, we find
that $\Delta R \sim R$, i.e. the outflow expands to twice the initial
radius by the time the induced shockwave reflects back to the edge of
the disc. At the edge, the shockwave now encounters outflow pressure
which is a factor $\sim 4$ lower than the pressure which drove the
shockwave. Assuming that the shockwave did not lose a significant
fraction of its energy (and pressure) during its vertical passage
through the disc, it causes the disc to expand until the pressure
approximately equilibrates. The expansion factor in disc thickness is
comparable to the pressure ratio, i.e. $H'/H \sim 4$. Given that the
external pressure becomes progressively more dominant over disc
internal pressure as the outflow expands (see
Fig. \ref{fig:fiducial_burstmax} and Section \ref{sec:outskirts}),
disc expansion should be more prominent in the outer parts of the
disc, leading to flaring qualitatively consistent with observations
\citep{vanderKruit1981A&A, Momany2006A&A, Merrett2006MNRAS,
  Vollmer2016A&A}.

\subsection{Comparison with other work}

In recent years, several authors investigated AGN-triggered star
formation in galaxy discs. \citet{Silk2013ApJ} provides estimates of
the star formation rate enhancement due to increased pressure, with
general results very similar to ours. The conclusions of that paper
state that quasar-mode feedback should be able to produce such
increased pressures, although the focus of the paper was on
jet-induced pressure increase. We show that high pressures, similar to
the fiducial values assumed by \citet{Silk2013ApJ}, are a plausible
result of AGN wind-driven outflows.

Numerical simulations have so far considered mainly compression of
galaxy discs by AGN jet-inflated bubbles \citep{Gaibler2012MNRAS,
  Bieri2016MNRAS}. Although these simulations consider significantly
different physical processes from ours, the general results are
qualitatively similar. In particular, \citet{Gaibler2012MNRAS} found
that a jet-driven outflow in a clumpy medium produces a bubble which
expands and compresses a turbulent galaxy disc, leading to enhancement
of the star formation rate by a factor of a few from the baseline
undisturbed disc SFR. \citet{Bieri2016MNRAS} found a stronger SFR
enhancement, but investigated a longer-duration evolution of a model
system, in which a jet-driven cocoon cools down and falls back on to
the galaxy disc. Our results, therefore, provide a somewhat different
mechanism for inducing higher rates of star formation in a galaxy
disc, one which operates on timescales of 1-100 Myr. However, more
detailed simulations are required in order to distinguish the
predictions of this model from those mentioned above.

\section{Summary and conclusion}\label{sec:concl}

We have used a largely analytic model to follow the evolution of a
spherically--symmetric, energy-driven AGN outflow expanding in a
realistic galaxy environment with an NFW halo and a Hernquist
bulge. We showed that if that the nucleus remains active
only long enough to inject enough energy into the gas to unbind it,
then the outflow can only be detected simultaneously with the AGN
when its radius is $\lesssim 10$~kpc, consistent
with observations. Outflows detected at greater distances in inactive
galaxies may point to recent episodes of nuclear activity which have
already ended. We also predict that outflows in more massive active
galaxies will typically be detected at larger distances from the
nucleus than those in lower--mass galaxies.

We have shown that the outflow pressure can enhance star formation in
a galaxy disc, producing observable differences in the normalisation
of the KS law for the host, at least if the disc is not extremely
gas--rich. The SFR enhancement is strongest in the centre and the
outskirts of the disc ($<1$~kpc and $>10$~kpc from the centre,
respectively), so these regions are the best places to look for the
effects of recent AGN activity. In particular, enhanced star formation
efficiency in the outskirts of a galaxy's disc can be used as another
diagnostic of an AGN phase within the last few times $10^7$~yr.

\section*{Acknowledgments}

KZ is funded by the Research Council Lithuania through the National
Science Programme grant no. LAT-09/2016. Astrophysics research at the
University of Leicester is funded by an STFC Consolidated grant.

\end{document}